\documentclass[letters,fleqn,usenatbib]{mnras}

\usepackage[T1]{fontenc}
\usepackage{ae,aecompl}

\usepackage{graphicx}	
\usepackage{amsmath}	
\usepackage{amssymb}	

\newcommand{\lppr}{\stackrel{<}{\scriptstyle \sim}}
\newcommand{\lappr}{\raisebox{-0.4ex}{$\lppr$}}

\title[The first post-merger ELM pre-WD in a wide binary.]{HE\,0430--2457: a post merger extremely low-mass pre-white dwarf in a wide binary posing as an extreme horizontal branch star}

\author[J. Vos et al.]{
Joris Vos$^{1}$\thanks{E-mail: joris.vos@uv.cl},
Monica Zorotovic$^{1}$,
Maja Vu\v{c}kovi\'{c}$^{1}$,
Matthias R. Schreiber$^{1,2}$,
\newauthor
Roy \O{}stensen$^{3}$
\\
$^{1}$Instituto de F\'{\i}sica y Astronom\'{\i}a, Universidad de Valparaiso, Gran Breta\~{n}a 1111, Playa Ancha, Valpara\'{\i}so 2360102, Chile\\
$^{2}$Nucleo Milenio on Planet Formation, Universidad Valparaiso, Avenida Gran Breta\~{n}a 1111, Valpara\'{\i}so
2360102, Chile\\
$^{3}$Department of Physics, Astronomy, and Materials Science, Missouri State University, Springfield, MO 65804, USA\\
}

\date{Accepted XXX. Received YYY; in original form ZZZ}

\pubyear{2017}

\begin{document}
\label{firstpage}
\pagerange{\pageref{firstpage}--\pageref{lastpage}}
\maketitle

\begin{abstract}
We report the discovery of HE\,0430--2457, the first extremely low-mass pre-white dwarf (ELM pre-WD) in a long period binary (P = 771 $\pm$ 3 d).
The spectroscopic parameters of the primary are determined to be T$_{\rm eff}$ = 26\,200 $\pm$ 1500 K and $\log{g} = 5.40 \pm 0.35$, placing it in the region occupied by core He-burning hot subdwarf B stars. By comparing the spectroscopic parameters of the K-type companion to stellar models, and using the mass ratio, the mass of the hot primary is determined to be 0.23 M$_{\odot}$. Given that this is too low for core He-burning, the primary in HE\,0430--2457 is not an EHB star but a pre WD of the ELM type.
As the lifetime of ELM pre-WDs in this region of the HR diagram populated by EHBs is thought to be very short, they are not considered to be part of the observed EHBs. However, the discovery of this system indicates that the percentage of ELM pre-WDs in the observed EHB population might be higher than previously thought.
Binary evolution models indicate that HE\,0430--2457 is likely formed by a merger of the inner binary in a hierarchical triple system.
\end{abstract}




\newcommand\kms{km s$^{-1}$}

\section{Introduction}
Binary interactions resulting in strong mass loss on the red giant branch (RGB) give birth to a variety of different objects residing on the extreme horizontal branch (EHB). This includes hot subdwarf B-type stars (sdBs) as well as extremely-low-mass white dwarfs (ELM WDs).

Both of these stellar types have lost most of their hydrogen envelope due to interaction with a binary companion while ascending the RGB. They differ in how far along the RGB they were when this interaction took place, leading to different core masses. \citet{Heber2016} defines sdB stars as core He-burning stars with a mass close to the canonical He-flash mass of 0.47 M$_{\odot}$. ELM WDs are stellar remnants that are too light to ignite He. They have lost their H-envelope on the Hertzsprung gap or early on the RGB and have masses $\leq$ 0.3 M$_{\odot}$ \citep{Brown2010}. 

sdBs and ELM WDs are usually differentiated spectroscopically, with sdBs having effective temperatures between 20\,000 and 35\,000 K and surface gravities of $5 < \log{g} < 6$, corresponding with the core He-burning EHB. ELM WDs have typical ranges in effective temperatures and surface gravities of 8\,000 $\leq T_{\rm eff} \leq$ 22\,000 K and $5 \leq \log{g} \leq 7$ \citep{Brown2013}, distinguishing them from core H-burning stars which have surface gravities of $\log{g} < 4.75$ at similar effective temperatures. 

The hotter ELM WDs (M = 0.21 -- 0.28 M$_{\odot}$, \citealt{Istrate2014}) can pass through the region of the HR diagram occupied by the sdB stars while transiting to the WD cooling track. While they cannot ignite core He-burning, they undergo phases of H-shell burning. The H-shell burning prolongs the transit time trough the sdB region of the HR diagram, but the transit times ($\leq$10 Myr, \citealt{Istrate2014}) are significantly shorter than the typical life time of a core He-burning sdB star ($\sim$100 Myr). 
During their transit from the RGB to the WD cooling track (before they switch from surface contraction to surface cooling), when they still have sufficient hydrogen in their shell to sustain H-shell burning, these stars are referred to as pre-WDs, proto-WDs or post-RGB stars. In this article the preposition `pre' is used to indicate that a star has not reached the WD cooling track. 
The first ELM pre-WD found with a surface gravity similar to that of an sdB star was HD\,188112 \citep{Heber2003}.

The main channels forming sdB stars are through common envelope (CE) ejection \citep{Webbink1984} leading to short period binaries, and stable Roche-lobe overflow (RLOF) resulting in wide binaries \citep{Han2002}. Both short period post-CE \citep[e.g.][]{Kupfer2015} and long period post-RLOF systems \citep[e.g.][]{Vos2017a} are observed. ELM WDs are thought to be formed by the CE evolution \citep{Marsh1995, Zorotovic2010}, which is supported by the observations of \citet{Brown2016}, who found that all currently known ELM WD binaries have periods of $\leq$ 1 day. To date no ELM WD was found in a wide binary.

HE\,0430--2457 is part of a long term observing program focusing on sdB+MS binaries. A spectral analysis places the primary with T$_{\rm eff}$ = 26200 $\pm$ 1500 K and $\log{g} = 5.40 \pm 0.35$ exactly where the core He-burning EHB stars are located \citep[Fig.\,7 in][]{Vos2018a}. The derived abundances of He, C, N, O, Mg, Si and Fe \citep[Table 7]{Vos2018a} match well with the average abundances of sdB stars given by \citet{Geier2013}. Spectroscopically, the primary component of HE\,0430--2457 looks exactly like an EHB star of the same temperature. However, HE\,0430--2457 is a wide binary with a K-type companion, and the derived mass ratio does not match with a 0.47 M${\odot}$ primary.
Given that the mass is not enough to sustain core He-burning, HE\,0430--2457 could only be an ELM pre-WD + K binary with a long orbital period, making it the first pre-WD of the ELM type in a wide binary.

At first glance this seems to contradict the CE origin suggested for ELM WDs. However, we argue that the ELM pre-WD in HE\,0430--2457 is most likely the result of a merger and that the progenitor of HE\,0430--2457 was a hierarchical triple system.
Hereafter we will refer to the hot ELM pre-WD star in HE\,0430--2457 as the WD component and to the cool companion as the K component.

\section{Radial velocities} \label{sect:rvs}

\begin{table}
 \centering
 \caption{Absolute parameters of HE\,0430--2457.}
 \label{tb:parameters}
 \begin{tabular}{lr@{ $\pm$ }lr@{ $\pm$ }l}
 \hline\hline
 \noalign{\smallskip}
 Parameter    &   \multicolumn{2}{c}{WD} & \multicolumn{2}{c}{K} \\\hline
 \noalign{\smallskip}
 $P$ (d)                              &    \multicolumn{2}{r@{ }}{771}     & \multicolumn{2}{@{$\pm$ }l}{3}   \\
 $T_0$ (d)                            &    \multicolumn{2}{r@{ }}{2456920} & \multicolumn{2}{@{$\pm$ }l}{5}   \\
 $q$ (WD/K)                           &    \multicolumn{2}{r@{ }}{0.33}    & \multicolumn{2}{@{$\pm$ }l}{0.05}\\
 $K$ (\kms)                  &  18.3     &  0.2   &  6.1      &  0.8       \\
 $\gamma$ (\kms)             &  6.7      &  0.2   &  4.5      &  0.5      \\
 T$_{\rm eff}$ (K)           &  26200    &  1500  &  4700     &  250      \\
 $\log{g}$ (dex)             &  5.40     &  0.35  &  4.70     &  0.40     \\
 L (L$_{\odot}$)             &  12       &  5     &  0.26     &  0.20     \\
 M (M$_{\odot}$)             &  0.23     &  0.05  &  0.71     &  0.09     \\
 R (R$_{\odot}$)             &  0.16     &  0.06  &  0.7      &  0.3      \\
 $[$Fe/H$]$ (dex)            & \multicolumn{2}{c}{/} &  -0.42 &  0.35     \\
 \hline
 \end{tabular}
\end{table}

\begin{figure}
    \includegraphics{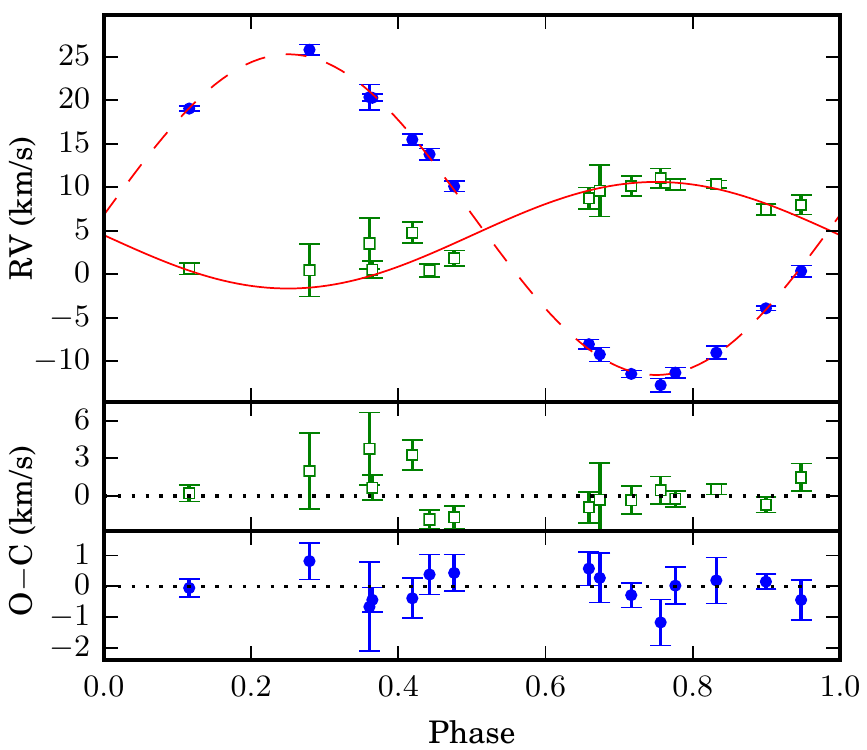}
    \caption{The phase folded RV curves for HE\,0430$-$2457. Top: orbital solution (solid line: K-star, dashed line: WD), and the observed RVs (open symbols: K-star, filled symbols: WD). Middle: residuals of the K-star. Bottom: residuals of the WD.}
    \label{fig:rv_fit}
\end{figure}

HE\,0430--2457 was on one of the selected candidates of the wide hot subdwarf binaries search programe as described in \citet{Vos2018a}. It has been observed with the UVES spectrograph attached to the 8.2\,m VLT telescope (UT2) on Cerro Paranal, Chile. UVES was used in standard dichroic-2 437+760 mode covering a wavelength range of 373 - 499\,nm in the BLUE and 565 - 946\,nm in the RED. The observations were taken with a slit of 1", reaching a resolution of 40\,000. In total 15 spectra with an average S/N of 40 at 5800 \AA\ were obtained between October 2011 and August 2017. 

To determine the radial velocities (RVs) we used a cross correlation (CC) with a template spectrum. For both the WD and the K-type companion we used a synthetic spectrum based on the best fitting parameters from \citet{Vos2018a} as a template. The WD has many sharp metal lines present in the spectra, and the RVs could be determined with high accuracy. The lines used in the CC are \ion{He}{i}, \ion{N}{ii}, \ion{C}{ii} and \ion{O}{ii}. The K-type companion is not very strong, and its lines are broadened due to its high rotational velocity ($V_{\rm rot} \sin{i}$ = 30 km s$^{-1}$, \citealt{Vos2018a}). In the CC we used the wavelength range 6000 - 6500 \AA\ while skipping the regions containing telluric lines or lines from the WD. The errors on the RVs are determined by using a Monte-Carlo method where in each iteration noise is added to the spectrum after which the CC is repeated. The error of the RV is the standard deviation of all derived RV measurements. A detailed description of the RV determination is given in \citep{Deca2018}.

The orbital parameters are determined by fitting a Keplerian curve to the observed RVs while varying the orbital period ($P$), time of periastron ($T_0$), eccentricity ($e$), angle of periastron ($\omega$), and amplitudes for both components ($K_{\rm{MS}}$ and $K_{\rm{sdB}}$) and systemic velocities ($\gamma_{\rm{MS}}$ and $\gamma_{\rm{sdB}}$). To check if the eccentricity is significant, the \citet{Lucy1971} test is used. It shows that the probability of falsely rejecting a circular orbit is P = 0.59. \citet{Lucy1971} argue that if P > 0.05, the orbit is effectively circular. In determining the final orbital parameters, the eccentricity is fixed at zero. The analysis of the RVs is the same as that used for composite sdB binaries in \citep{Vos2012, Vos2013, Vos2017a}. The RVs and the best fitting Keplerian curves are shown in Fig.\,\ref{fig:rv_fit}, while the resulting parameters and their errors are given in Table\,\ref{tb:parameters}.

\section{Spectral parameters}\label{sect:spectral_parameters}

\begin{figure}
    \includegraphics{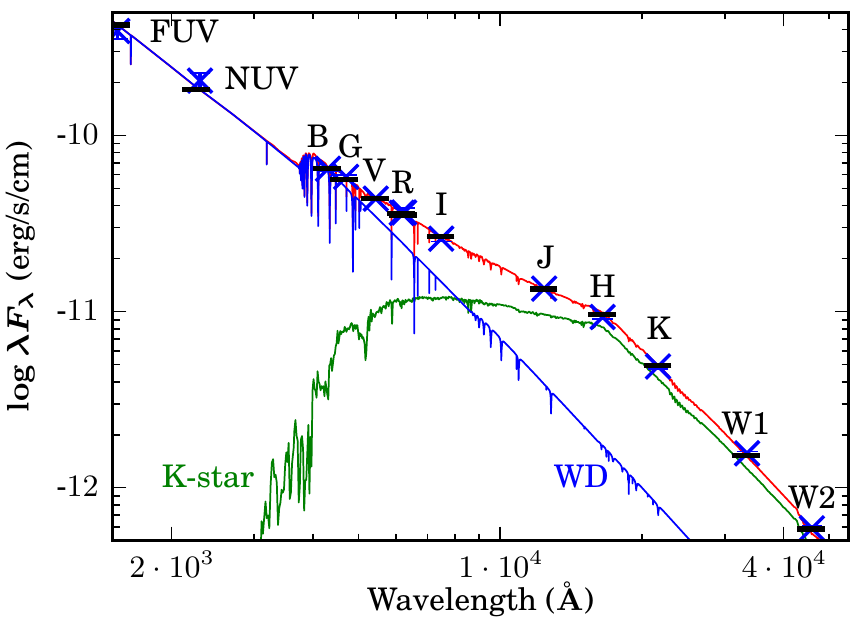}
    \caption{Photometric SED of HE\,0430$-$2457 with the best fitting model. The observed photometry is plotted in blue crosses, while the synthetic best fit photometry is plotted in black horizontal line. The best fitting binary model is shown with a red line, while the models for the K and WD components are shown respectively with a green and a blue line.}
    \label{fig:sed_fit}
\end{figure}

\citet{Vos2018a} performed a detailed spectral analysis of both the WD and K component of HE\,0430-2457. The {\sc XTgrid} code \citep{Nemeth2012} was used to derive spectral parameters for both the WD and K companion by simultaneously fitting both components in the highest S/N UVES spectrum. {\sc XTgrid} uses the NLTE model atmosphere code {\sc Tlusty} \citep{Hubeny1995} for the WD combined with the {\sc Phoenix} model atmosphere code (version 1, \citealt{Husser2013}) for the cool companion. A second set of spectral parameters for the K companion was derived using the Grid Search in Stellar Parameters (GSSP) code \citep{Tkachenko2015} on a master spectrum created by combining all UVES spectra shifted to the zero point velocity of the K star. GSSP uses the {\sc LLmodels} code \citep{Shulyak2004}. A detailed description of these methods is given in \citet{Vos2018a}.

A third set of spectral parameters is determined from the photometric spectral energy distribution (SED) using photometry from GALEX DR5 (\citealt{Bianchi2011} using the correction for bright targets of \citealt{Camarota2014}), APASS DR9 \citep{Henden2016}, 2MASS \citep{Skrutskie2006} and WISE \citep{Cutri2012}. The SED is fitted with TMAP (T\"{u}bingen NLTE Model-Atmosphere Package, \citealt{Werner2003}) atmosphere models for the WD and Kurucz atmosphere models \citep{Kurucz1979} for the K star. The fit uses a grid based approach, where 10\,000\,000 atmosphere models are randomly chosen in the full parameter space, and the best model is the one with the lowest $\chi^2$. The mass ratio of the system is used to provide an extra constraint in the fit. A detailed description of this approach is given in \citet{Degroote2011} and \citet{Vos2012, Vos2013}.

The SED fit results in T$_{\rm eff}$ = 25700 $\pm$ 2500 K and $\log{g}$ = 5.5 $\pm$ 0.5 dex for the WD, while for the cool companion we find T$_{\rm eff}$ = 4700 $\pm$ 300 K and $\log{g}$ = 4.6 $\pm$ 0.5 dex. The SED fitting procedure provides the luminosities of both components as respectively 12 $\pm$ 5 L$_{\odot}$ and 0.26 $\pm$ 0.20 L$_{\odot}$, as well as the distance to the system of d = 720 $\pm$ 80 pc. These results are in very good agreement with the results from {\sc XTgrid} and GSSP. The observed photometry together with the best fitting binary SED model is shown in Fig.\,\ref{fig:sed_fit}.

The final spectral parameters were determined by taking the average of those determined by the different methods weighted by their respective errors. In Table\,\ref{tb:parameters} all spectroscopic parameters are summarized.

\section{Mass determination}\label{sect:mass_determination}

\begin{figure}
    \includegraphics{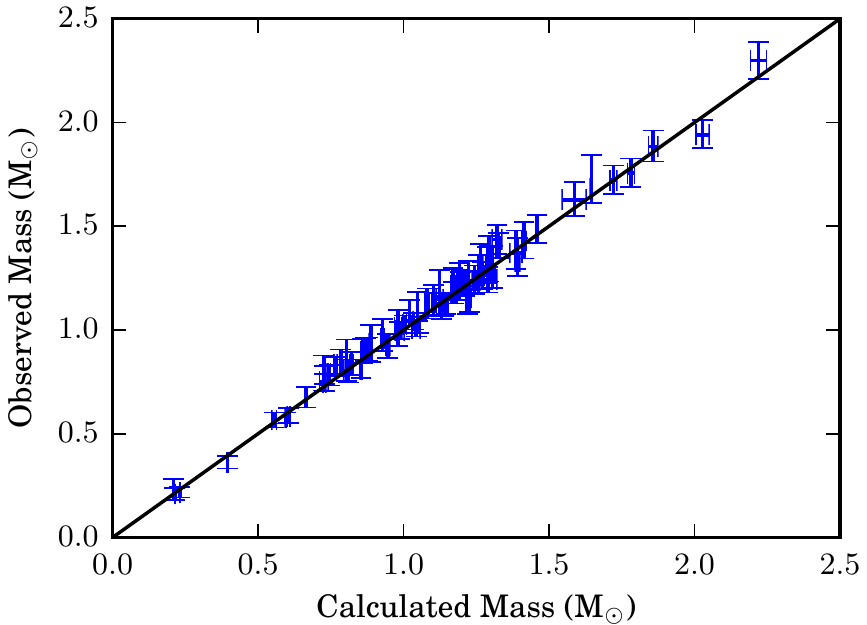}
    \caption{Comparison between the observed and calculated mass based on the spectroscopic parameters for 68 double lined eclipsing binaries taken from the DEBCat catalog.}
    \label{fig:emcmass_validation}
\end{figure}

By comparing the spectroscopic parameters of the K-type star with stellar evolution models we can calculate the posterior probability distribution of its mass based on the observed effective temperature, surface gravity, luminosity and metalicity. To achieve this we use a Markov-chain-Monte-Carlo (MCMC) approach similar to the one described in \citet{Maxted2015}. Our method differs in that it is implemented in python using the {\sc emcee} package \citep{Foreman-Mackey2013} using the surface gravity as an observable instead of the radius which is used in \citet{Maxted2015}. To calculate the posterior probabilities, 100 walkers are randomly initialized in the parameter space covered by the stellar evolution models. Each walker is allowed to take 2000 steps, of which the first 200 are removed to let the walker settle on its fit (burn in). The remaining steps are used to calculate the posterior distribution.

This code uses the Yale - Potsdam stellar isochrone (YaPSI, \citealt{Spada2017}) models\footnote{http://www.astro.yale.edu/yapsi/}. This grid of stellar evolution models is an update of the Yonsei-Yale stellar evolution models constructed with the purpose to derive accurate stellar parameters of exoplanet host stars. This grid has a range in mass of 0.15 - 5.0 M$_{\odot}$, in age from 1 - 20 Gyr and is calculated for metalicities ranging from -1.5 to +0.3.

We used the catalogue of the physical properties of eclipsing binaries (DEBCat, \citealt{Southworth2015} based on \citealt{Andersen1991}) to test our implementation of the MCMC algorithm. DEBCat contains 68 MS stars with known masses, effective temperatures, surface gravities, luminosities and metalicities in the range of the evolution models. For each system we have calculated the mass based on the spectral parameters. In Fig.\,\ref{fig:emcmass_validation} the observed mass is plotted versus the mass calculated from the spectroscopic parameters using our MCMC method. For all systems, the calculated mass corresponds very well with the derived mass. The average absolute difference between the observed and calculated mass is 0.03 M$_{\odot}$, while the maximum difference is 0.11 M$_{\odot}$. Furthermore, for each system, the calculated mass fits with the observed mass within the errors on the calculated mass.

As a comparison, the same calculation was performed using the MESA Isochrones and Stellar Tracks
(MIST, \citealt{Dotter2016, Choi2016}) instead of the YaPSI models. These models are calculated using the MESA stellar evolution code \citep[][r7503]{Paxton2011, Paxton2013, Paxton2015}. The mass and metalicity range of the MIST models is wider than for YaPSI, but the region of interest for our case is equally well sampled. The results of the YaPSI models for the eclipsing-binary sample are marginally better, but the difference between the two sets of models is negligible. To derive the mass of HE\,0430$-$2457, the YaPSI models were used.

Using this method we find a mass of 0.71 $\pm$ 0.09 M$_{\odot}$ for the K-type companion of HE\,0430--2457. Using the mass ratio derived from the RV curve, we find a mass of 0.23 $\pm$ 0.05 M$_{\odot}$ for the WD, placing it firmly in the mass range for ELM WDs. Based on this mass and surface gravity derived from the spectral analysis, the radii of both components are calculated to be 0.16 $\pm$ 0.06 R$_{\odot}$ and 0.7 $\pm$ 0.3 R$_{\odot}$ for respectively the WD and the K component, and the inclination of the orbit is $i = 21 \pm 7$ dgr.

If the WD would be a core He-burning sdB star with a canonical mass of 0.47 M$_{\odot}$, the K-type companion would have a mass of 1.42 $\pm$ 0.23 M$_{\odot}$. For a star with this mass to have the observed temperature, it would have to be a subgiant. However, the YaPSI models show that a 1.42 M$_{\odot}$ star with the observed metalicity and effective temperature would have a radius of $\sim$ 15 R$_{\odot}$. Such a star would completely dominate the SED at the visible and red wavelengths, which is obviously not the case. Furthermore, the V$_{\rm rot}\,\sin{i}$ of 30 km\,s$^{-1}$ would be unreasonably high for a subgiant.
The lowest mass for a star to still sustain core He-burning is 0.33 M$_{\odot}$, if it ignited He in non-degenerate conditions. If the WD would have this mass, the K-companion would have a mass of 1.0 $\pm$ 0.1 M$_{\odot}$. Following the same reasoning as before a subgiant would have a radius of $\sim$ 10 R$_{\odot}$, which can be rejected based on the SED. We thus conclude that the primary of HE\,0430-2457 is indeed an ELM pre-WD.

\section{Formation history}\label{sect:formation_history}

Although the MS life time of the progenitors of low-mass (LM) WDs (M$_{\rm WD} \lesssim$ 0.5 M$_{\odot}$) largely exceeds the Hubble time, they make up roughly 10 per cent of the observed WDs in the solar neighborhood. Their existence is most likely explained by binary star interactions, in particular CE evolution as outlined in the introduction. Indeed, most of the observed LM WDs are members of close binary systems \citep{Rebassa-Mansergas2011} and until our discovery of HE\,0430$-$2457, the observed close binary fraction of ELM WDs was 100 per cent. However, 20-30 per cent of the known LM WDs are apparently single stars \citep{Brown2011}.  

Recently, \citet{Zorotovic2017} showed that the existence of single LM WDs can be naturally explained in the context of the new model for the evolution of cataclysmic variables proposed by \citet{Schreiber2016}, in which systems with LM WDs are more likely to merge after mass transfer starts. We propose a similar solution for HE\,0430$-$2457, where the ELM pre-WD observed is the result of the merger of a close (inner) binary in a hierarchical triple system with a wide K companion that we still observe today.

We used the Binary Star Evolution (BSE) code \citep{Hurley2002} to perform simulations of binary star populations and to identify the conditions for the merging of an ELM WD with a LM companion. We found possible solutions if the inner binary is composed of an 1.4 - 2.3 M$_{\odot}$ ELM WD progenitor with a 0.38 - 0.65 M$_{\odot}$ companion at a very short orbital period of $\sim$ 1 - 3 days.
These systems may evolve through a phase of stable non-conservative mass transfer which ends when the primary has lost its entire envelope. The emerging close detached binary consists of an ELM pre-WD and a LM MS companion (0.4 - 0.9 M$_{\odot}$) with an orbital period similar to the initial one (a few days). The two stars are then brought closer together by angular momentum loss through gravitational radiation and especially magnetic braking. When the MS star fills its Roche-lobe, dynamically unstable mass transfer is started, developing a CE-like event where a fraction of the companion mass is expelled due to dynamical friction during the spiraling-in process. The remaining H-rich material is accreted by the WD and is likely to cause H-shell burning, converting the WD into a giant-like star with a degenerate core. 
The He produced by the H-shell burning may slightly increase the mass of the finally emerging single ELM WD.

The details of the outlined merger process 
are very complicated and not well understood, but it is so far the only physically plausible explanation for the existence of this star without a close companion. The fact that the final mass of the emerging WD is still in the ELM regime ($\lappr\,0.3$ M$_{\odot}$) implies that mass growth of the He-core during the merger process must have been very small. This could be due to efficient use of the available orbital energy (of the initial close binary) at the onset of the merger process to expel large parts of the secondary star in the CE-like event, and/or strong mass loss of the giant-like star before the core grows substantially. HE\,0430$-$2457 can therefore be used to derive constraints on the merger theory of cataclysmic variables.

If the evolutionary path that we propose here is correct, the K-type star companion we observe today in a wide orbit must have been initially a third object in a hierarchical triple, with an initial period of roughly 100 days, that migrated further out as a consequence of mass loss in the inner binary. Interestingly, while the range of possible initial masses is relatively broad, the short initial period of the inner binary of 1-3 days is a strict requirement for producing a single ELM WD due to binary star evolution. Exactly these type of close MS binaries are found to nearly always have a third companion \citep{Tokovinin2006}. 
The fact that the ELM pre-WD in HE\,0430$-$2457 has a wide companion is therefore perfectly consistent with the evolutionary pathway we suggest.

\section{Conclusions}\label{sect:discussion}
Compared to the typical parameters of ELM WDs, HE\,0430--2457 has a higher T$_{\rm eff}$ of 26\,200 K, placing it in the T$_{\rm eff}$ -- $\log{g}$ region that is normally occupied by core He-burning EHB stars.

The transit time of ELM WDs between the RGB and the WD cooling sequence depends on their final mass. Those systems with a high enough mass to reach a T$_{\rm eff}$ similar to sdB stars spend between 1 and 10 Myr in the region of the HR diagram occupied by the sdBs. Due to these short transit times, ELM WDs are usually not considered to be part of the observed sdB population. The observation of an ELM pre-WD in this this T$_{\rm eff}$ -- $\log{g}$ range can have important consequences for our understanding of sdB stars.

Currently it is assumed that sdB stars have a mass close to the canonical mass of 0.47 M$_{\odot}$. If the actual mass range would deviate significantly, properties derived for companion stars based on this assumption might be significantly off. An example for this would be the differentiation between brown dwarf and WD companions of close sdB binaries \citep[e.g.][]{Kupfer2015}.

Binary population synthesis studies focusing on both post-CE and post-RLOF sdB binaries are validated by comparing their theoretical populations to the observed population assuming that the observed population is homogeneous. Deviations from this assumption would impact the conclusions of such studies.

Finally, the evolutionary history we suggest for HE\,0430-2457, i.e. the merger of the inner binary in a hierarchical triple system, represents a hidden potential of WD+MS mergers as explanation of single LM WDs or LM WDs in wide binaries.

\vspace{-0.3cm}

\section*{Acknowledgements}
Based on observations collected at the European Organization for Astronomical Research in the Southern Hemisphere under ESO programmes 088.D-0364(A), 093.D-0629(A), 096.D-0180(A), 097.D-0110(A), 098.D-0018(A), 099.D-0014(A) and 0100.D-0082(A).
JV, MZ and MRS acknowledge financial support from FONDECYT grants number 3160504, 11170559 and 1181404, respectively. 



\bibliographystyle{mnras}
\bibliography{bibliogaphy} 

\begin{thebibliography}{}
\makeatletter
\relax
\def\mn@urlcharsother{\let\do\@makeother \do\$\do\&\do\#\do\^\do\_\do\%\do\~}
\def\mn@doi{\begingroup\mn@urlcharsother \@ifnextchar [ {\mn@doi@}
  {\mn@doi@[]}}
\def\mn@doi@[#1]#2{\def\@tempa{#1}\ifx\@tempa\@empty \href
  {http://dx.doi.org/#2} {doi:#2}\else \href {http://dx.doi.org/#2} {#1}\fi
  \endgroup}
\def\mn@eprint#1#2{\mn@eprint@#1:#2::\@nil}
\def\mn@eprint@arXiv#1{\href {http://arxiv.org/abs/#1} {{\tt arXiv:#1}}}
\def\mn@eprint@dblp#1{\href {http://dblp.uni-trier.de/rec/bibtex/#1.xml}
  {dblp:#1}}
\def\mn@eprint@#1:#2:#3:#4\@nil{\def\@tempa {#1}\def\@tempb {#2}\def\@tempc
  {#3}\ifx \@tempc \@empty \let \@tempc \@tempb \let \@tempb \@tempa \fi \ifx
  \@tempb \@empty \def\@tempb {arXiv}\fi \@ifundefined
  {mn@eprint@\@tempb}{\@tempb:\@tempc}{\expandafter \expandafter \csname
  mn@eprint@\@tempb\endcsname \expandafter{\@tempc}}}

\bibitem[\protect\citeauthoryear{{Andersen}}{{Andersen}}{1991}]{Andersen1991}
{Andersen} J.,  1991, \mn@doi [\aapr] {10.1007/BF00873538}, \href
  {http://adsabs.harvard.edu/abs/1991A%26ARv...3...91A} {3, 91}

\bibitem[\protect\citeauthoryear{{Bianchi}, {Herald}, {Efremova}, {Girardi},
  {Zabot}, {Marigo}, {Conti}  \& {Shiao}}{{Bianchi} et~al.}{2011}]{Bianchi2011}
{Bianchi} L.,  {Herald} J.,  {Efremova} B.,  {Girardi} L.,  {Zabot} A.,
  {Marigo} P.,  {Conti} A.,   {Shiao} B.,  2011, \mn@doi [\apss]
  {10.1007/s10509-010-0581-x}, \href
  {http://adsabs.harvard.edu/abs/2011Ap%26SS.335..161B} {335, 161}

\bibitem[\protect\citeauthoryear{{Brown}, {Kilic}, {Allende Prieto}  \&
  {Kenyon}}{{Brown} et~al.}{2010}]{Brown2010}
{Brown} W.~R.,  {Kilic} M.,  {Allende Prieto} C.,   {Kenyon} S.~J.,  2010,
  \mn@doi [\apj] {10.1088/0004-637X/723/2/1072}, \href
  {http://adsabs.harvard.edu/abs/2010ApJ...723.1072B} {723, 1072}

\bibitem[\protect\citeauthoryear{{Brown}, {Kilic}, {Allende Prieto}  \&
  {Kenyon}}{{Brown} et~al.}{2011}]{Brown2011}
{Brown} W.~R.,  {Kilic} M.,  {Allende Prieto} C.,   {Kenyon} S.~J.,  2011,
  \mn@doi [\mnras] {10.1111/j.1745-3933.2010.00986.x}, \href
  {http://adsabs.harvard.edu/abs/2011MNRAS.411L..31B} {411, L31}

\bibitem[\protect\citeauthoryear{{Brown}, {Kilic}, {Allende Prieto},
  {Gianninas}  \& {Kenyon}}{{Brown} et~al.}{2013}]{Brown2013}
{Brown} W.~R.,  {Kilic} M.,  {Allende Prieto} C.,  {Gianninas} A.,   {Kenyon}
  S.~J.,  2013, \mn@doi [\apj] {10.1088/0004-637X/769/1/66}, \href
  {http://adsabs.harvard.edu/abs/2013ApJ...769...66B} {769, 66}

\bibitem[\protect\citeauthoryear{{Brown}, {Gianninas}, {Kilic}, {Kenyon}  \&
  {Allende Prieto}}{{Brown} et~al.}{2016}]{Brown2016}
{Brown} W.~R.,  {Gianninas} A.,  {Kilic} M.,  {Kenyon} S.~J.,   {Allende
  Prieto} C.,  2016, \mn@doi [\apj] {10.3847/0004-637X/818/2/155}, \href
  {http://adsabs.harvard.edu/abs/2016ApJ...818..155B} {818, 155}

\bibitem[\protect\citeauthoryear{{Camarota} \& {Holberg}}{{Camarota} \&
  {Holberg}}{2014}]{Camarota2014}
{Camarota} L.,  {Holberg} J.~B.,  2014, \mn@doi [\mnras]
  {10.1093/mnras/stt2422}, \href
  {http://adsabs.harvard.edu/abs/2014MNRAS.438.3111C} {438, 3111}

\bibitem[\protect\citeauthoryear{{Choi}, {Dotter}, {Conroy}, {Cantiello},
  {Paxton}  \& {Johnson}}{{Choi} et~al.}{2016}]{Choi2016}
{Choi} J.,  {Dotter} A.,  {Conroy} C.,  {Cantiello} M.,  {Paxton} B.,
  {Johnson} B.~D.,  2016, \mn@doi [\apj] {10.3847/0004-637X/823/2/102}, \href
  {http://adsabs.harvard.edu/abs/2016ApJ...823..102C} {823, 102}

\bibitem[\protect\citeauthoryear{{Cutri} \& {et al.}}{{Cutri} \& {et
  al.}}{2012}]{Cutri2012}
{Cutri} R.~M.,  {et al.} 2012, VizieR Online Data Catalog, \href
  {http://cdsads.u-strasbg.fr/abs/2012yCat.2311....0C} {2311}

\bibitem[\protect\citeauthoryear{{Deca}, {Vos}, {N{\'e}meth}, {Maxted},
  {Copperwheat}, {Marsh}  \& {{\O}stensen}}{{Deca} et~al.}{2018}]{Deca2018}
{Deca} J.,  {Vos} J.,  {N{\'e}meth} P.,  {Maxted} P.~F.~L.,  {Copperwheat}
  C.~M.,  {Marsh} T.~R.,   {{\O}stensen} R.,  2018, \mn@doi [\mnras]
  {10.1093/mnras/stx2755}, \href
  {http://adsabs.harvard.edu/abs/2018MNRAS.474..433D} {474, 433}

\bibitem[\protect\citeauthoryear{{Degroote} et~al.,}{{Degroote}
  et~al.}{2011}]{Degroote2011}
{Degroote} P.,  et~al., 2011, \mn@doi [\aap] {10.1051/0004-6361/201116802},
  \href {http://adsabs.harvard.edu/abs/2011A%26A...536A..82D} {536, A82}

\bibitem[\protect\citeauthoryear{{Dotter}}{{Dotter}}{2016}]{Dotter2016}
{Dotter} A.,  2016, \mn@doi [\apjs] {10.3847/0067-0049/222/1/8}, \href
  {http://adsabs.harvard.edu/abs/2016ApJS..222....8D} {222, 8}

\bibitem[\protect\citeauthoryear{{Foreman-Mackey}, {Hogg}, {Lang}  \&
  {Goodman}}{{Foreman-Mackey} et~al.}{2013}]{Foreman-Mackey2013}
{Foreman-Mackey} D.,  {Hogg} D.~W.,  {Lang} D.,   {Goodman} J.,  2013, \mn@doi
  [\pasp] {10.1086/670067}, \href
  {http://adsabs.harvard.edu/abs/2013PASP..125..306F} {125, 306}

\bibitem[\protect\citeauthoryear{{Geier}}{{Geier}}{2013}]{Geier2013}
{Geier} S.,  2013, \mn@doi [\aap] {10.1051/0004-6361/201220549}, \href
  {http://adsabs.harvard.edu/abs/2013A%26A...549A.110G} {549, A110}

\bibitem[\protect\citeauthoryear{{Han}, {Podsiadlowski}, {Maxted}, {Marsh}  \&
  {Ivanova}}{{Han} et~al.}{2002}]{Han2002}
{Han} Z.,  {Podsiadlowski} P.,  {Maxted} P.~F.~L.,  {Marsh} T.~R.,   {Ivanova}
  N.,  2002, \mn@doi [\mnras] {10.1046/j.1365-8711.2002.05752.x}, \href
  {http://adsabs.harvard.edu/abs/2002MNRAS.336..449H} {336, 449}

\bibitem[\protect\citeauthoryear{{Heber}}{{Heber}}{2016}]{Heber2016}
{Heber} U.,  2016, \mn@doi [\pasp] {10.1088/1538-3873/128/966/082001}, \href
  {http://adsabs.harvard.edu/abs/2016PASP..128h2001H} {128, 082001}

\bibitem[\protect\citeauthoryear{{Heber}, {Edelmann}, {Lisker}  \&
  {Napiwotzki}}{{Heber} et~al.}{2003}]{Heber2003}
{Heber} U.,  {Edelmann} H.,  {Lisker} T.,   {Napiwotzki} R.,  2003, \mn@doi
  [\aap] {10.1051/0004-6361:20031553}, \href
  {http://adsabs.harvard.edu/abs/2003A%26A...411L.477H} {411, L477}

\bibitem[\protect\citeauthoryear{{Henden}, {Templeton}, {Terrell}, {Smith},
  {Levine}  \& {Welch}}{{Henden} et~al.}{2016}]{Henden2016}
{Henden} A.~A.,  {Templeton} M.,  {Terrell} D.,  {Smith} T.~C.,  {Levine} S.,
  {Welch} D.,  2016, VizieR Online Data Catalog, \href
  {http://esoads.eso.org/abs/2016yCat.2336....0H} {2336}

\bibitem[\protect\citeauthoryear{{Hubeny} \& {Lanz}}{{Hubeny} \&
  {Lanz}}{1995}]{Hubeny1995}
{Hubeny} I.,  {Lanz} T.,  1995, \mn@doi [\apj] {10.1086/175226}, \href
  {http://adsabs.harvard.edu/abs/1995ApJ...439..875H} {439, 875}

\bibitem[\protect\citeauthoryear{{Hurley}, {Tout}  \& {Pols}}{{Hurley}
  et~al.}{2002}]{Hurley2002}
{Hurley} J.~R.,  {Tout} C.~A.,   {Pols} O.~R.,  2002, \mnras, \href
  {http://adsabs.harvard.edu/cgi-bin/nph-bib_query?bibcode=2002MNRAS.329..897H&db_key=AST}
  {329, 897}

\bibitem[\protect\citeauthoryear{{Husser}, {Wende-von Berg}, {Dreizler},
  {Homeier}, {Reiners}, {Barman}  \& {Hauschildt}}{{Husser}
  et~al.}{2013}]{Husser2013}
{Husser} T.-O.,  {Wende-von Berg} S.,  {Dreizler} S.,  {Homeier} D.,  {Reiners}
  A.,  {Barman} T.,   {Hauschildt} P.~H.,  2013, \mn@doi [\aap]
  {10.1051/0004-6361/201219058}, \href
  {http://adsabs.harvard.edu/abs/2013A%26A...553A...6H} {553, A6}

\bibitem[\protect\citeauthoryear{{Istrate}, {Tauris}, {Langer}  \&
  {Antoniadis}}{{Istrate} et~al.}{2014}]{Istrate2014}
{Istrate} A.~G.,  {Tauris} T.~M.,  {Langer} N.,   {Antoniadis} J.,  2014,
  \mn@doi [\aap] {10.1051/0004-6361/201424681}, \href
  {http://adsabs.harvard.edu/abs/2014A%26A...571L...3I} {571, L3}

\bibitem[\protect\citeauthoryear{{Kupfer} et~al.,}{{Kupfer}
  et~al.}{2015}]{Kupfer2015}
{Kupfer} T.,  et~al., 2015, VizieR Online Data Catalog, \href
  {http://adsabs.harvard.edu/abs/2015yCat..35760044K} {357}

\bibitem[\protect\citeauthoryear{{Kurucz}}{{Kurucz}}{1979}]{Kurucz1979}
{Kurucz} R.~L.,  1979, \mn@doi [\apjs] {10.1086/190589}, \href
  {http://cdsads.u-strasbg.fr/abs/1979ApJS...40....1K} {40, 1}

\bibitem[\protect\citeauthoryear{{Lucy} \& {Sweeney}}{{Lucy} \&
  {Sweeney}}{1971}]{Lucy1971}
{Lucy} L.~B.,  {Sweeney} M.~A.,  1971, \mn@doi [\aj] {10.1086/111159}, \href
  {http://adsabs.harvard.edu/abs/1971AJ.....76..544L} {76, 544}

\bibitem[\protect\citeauthoryear{{Marsh}, {Dhillon}  \& {Duck}}{{Marsh}
  et~al.}{1995}]{Marsh1995}
{Marsh} T.~R.,  {Dhillon} V.~S.,   {Duck} S.~R.,  1995, \mn@doi [\mnras]
  {10.1093/mnras/275.3.828}, \href
  {http://adsabs.harvard.edu/abs/1995MNRAS.275..828M} {275, 828}

\bibitem[\protect\citeauthoryear{{Maxted}, {Serenelli}  \&
  {Southworth}}{{Maxted} et~al.}{2015}]{Maxted2015}
{Maxted} P.~F.~L.,  {Serenelli} A.~M.,   {Southworth} J.,  2015, \mn@doi [\aap]
  {10.1051/0004-6361/201425331}, \href
  {http://adsabs.harvard.edu/abs/2015A%26A...575A..36M} {575, A36}

\bibitem[\protect\citeauthoryear{{N{\'e}meth}, {Kawka}  \&
  {Vennes}}{{N{\'e}meth} et~al.}{2012}]{Nemeth2012}
{N{\'e}meth} P.,  {Kawka} A.,   {Vennes} S.,  2012, \mn@doi [\mnras]
  {10.1111/j.1365-2966.2012.22009.x}, \href
  {http://adsabs.harvard.edu/abs/2012MNRAS.427.2180N} {427, 2180}

\bibitem[\protect\citeauthoryear{{Paxton}, {Bildsten}, {Dotter}, {Herwig},
  {Lesaffre}  \& {Timmes}}{{Paxton} et~al.}{2011}]{Paxton2011}
{Paxton} B.,  {Bildsten} L.,  {Dotter} A.,  {Herwig} F.,  {Lesaffre} P.,
  {Timmes} F.,  2011, \mn@doi [\apjs] {10.1088/0067-0049/192/1/3}, \href
  {http://adsabs.harvard.edu/abs/2011ApJS..192....3P} {192, 3}

\bibitem[\protect\citeauthoryear{{Paxton} et~al.,}{{Paxton}
  et~al.}{2013}]{Paxton2013}
{Paxton} B.,  et~al., 2013, \mn@doi [\apjs] {10.1088/0067-0049/208/1/4}, \href
  {http://adsabs.harvard.edu/abs/2013ApJS..208....4P} {208, 4}

\bibitem[\protect\citeauthoryear{{Paxton} et~al.,}{{Paxton}
  et~al.}{2015}]{Paxton2015}
{Paxton} B.,  et~al., 2015, \mn@doi [\apjs] {10.1088/0067-0049/220/1/15}, \href
  {http://adsabs.harvard.edu/abs/2015ApJS..220...15P} {220, 15}

\bibitem[\protect\citeauthoryear{{Rebassa-Mansergas}, {Nebot
  G{\'o}mez-Mor{\'a}n}, {Schreiber}, {Girven}  \&
  {G{\"a}nsicke}}{{Rebassa-Mansergas} et~al.}{2011}]{Rebassa-Mansergas2011}
{Rebassa-Mansergas} A.,  {Nebot G{\'o}mez-Mor{\'a}n} A.,  {Schreiber} M.~R.,
  {Girven} J.,   {G{\"a}nsicke} B.~T.,  2011, \mn@doi [\mnras]
  {10.1111/j.1365-2966.2011.18200.x}, \href
  {http://adsabs.harvard.edu/abs/2011MNRAS.413.1121R} {413, 1121}

\bibitem[\protect\citeauthoryear{{Schreiber}, {Zorotovic}  \&
  {Wijnen}}{{Schreiber} et~al.}{2016}]{Schreiber2016}
{Schreiber} M.~R.,  {Zorotovic} M.,   {Wijnen} T.~P.~G.,  2016, \mn@doi
  [\mnras] {10.1093/mnrasl/slv144}, \href
  {http://adsabs.harvard.edu/abs/2016MNRAS.455L..16S} {455, L16}

\bibitem[\protect\citeauthoryear{{Shulyak}, {Tsymbal}, {Ryabchikova},
  {St{\"u}tz}  \& {Weiss}}{{Shulyak} et~al.}{2004}]{Shulyak2004}
{Shulyak} D.,  {Tsymbal} V.,  {Ryabchikova} T.,  {St{\"u}tz} C.,   {Weiss}
  W.~W.,  2004, \mn@doi [\aap] {10.1051/0004-6361:20034169}, \href
  {http://adsabs.harvard.edu/abs/2004A%26A...428..993S} {428, 993}

\bibitem[\protect\citeauthoryear{{Skrutskie} et~al.,}{{Skrutskie}
  et~al.}{2006}]{Skrutskie2006}
{Skrutskie} M.~F.,  et~al., 2006, \mn@doi [\aj] {10.1086/498708}, \href
  {http://esoads.eso.org/abs/2006AJ....131.1163S} {131, 1163}

\bibitem[\protect\citeauthoryear{{Southworth}}{{Southworth}}{2015}]{Southworth2015}
{Southworth} J.,  2015, in {Rucinski} S.~M.,  {Torres} G.,   {Zejda} M.,  eds,
  ASPCS Vol. 496, Living Together: Planets, Host Stars and Binaries. p.~164
  (\mn@eprint {arXiv} {1411.1219})

\bibitem[\protect\citeauthoryear{{Spada}, {Demarque}, {Kim}, {Boyajian}  \&
  {Brewer}}{{Spada} et~al.}{2017}]{Spada2017}
{Spada} F.,  {Demarque} P.,  {Kim} Y.-C.,  {Boyajian} T.~S.,   {Brewer} J.~M.,
  2017, \mn@doi [\apj] {10.3847/1538-4357/aa661d}, \href
  {http://adsabs.harvard.edu/abs/2017ApJ...838..161S} {838, 161}

\bibitem[\protect\citeauthoryear{{Tkachenko}}{{Tkachenko}}{2015}]{Tkachenko2015}
{Tkachenko} A.,  2015, \mn@doi [\aap] {10.1051/0004-6361/201526513}, \href
  {http://adsabs.harvard.edu/abs/2015A%26A...581A.129T} {581, A129}

\bibitem[\protect\citeauthoryear{{Tokovinin}, {Thomas}, {Sterzik}  \&
  {Udry}}{{Tokovinin} et~al.}{2006}]{Tokovinin2006}
{Tokovinin} A.,  {Thomas} S.,  {Sterzik} M.,   {Udry} S.,  2006, \mn@doi [\aap]
  {10.1051/0004-6361:20054427}, \href
  {http://adsabs.harvard.edu/abs/2006A%26A...450..681T} {450, 681}

\bibitem[\protect\citeauthoryear{{Vos} et~al.,}{{Vos} et~al.}{2012}]{Vos2012}
{Vos} J.,  et~al., 2012, \mn@doi [\aap] {10.1051/0004-6361/201219723}, \href
  {http://adsabs.harvard.edu/abs/2012A%26A...548A...6V} {548, A6, Paper I}

\bibitem[\protect\citeauthoryear{{Vos}, {{\O}stensen}, {N{\'e}meth}, {Green},
  {Heber}  \& {Van Winckel}}{{Vos} et~al.}{2013}]{Vos2013}
{Vos} J.,  {{\O}stensen} R.~H.,  {N{\'e}meth} P.,  {Green} E.~M.,  {Heber} U.,
   {Van Winckel} H.,  2013, \mn@doi [\aap] {10.1051/0004-6361/201322200}, \href
  {http://adsabs.harvard.edu/abs/2013A%26A...559A..54V} {559, A54, Paper II}

\bibitem[\protect\citeauthoryear{{Vos}, {{\O}stensen}, {Vu{\v c}kovi{\'c}}  \&
  {Van Winckel}}{{Vos} et~al.}{2017}]{Vos2017a}
{Vos} J.,  {{\O}stensen} R.~H.,  {Vu{\v c}kovi{\'c}} M.,   {Van Winckel} H.,
  2017, \mn@doi [\aap] {10.1051/0004-6361/201730958}, \href
  {http://adsabs.harvard.edu/abs/2017A%26A...605A.109V} {605, A109}

\bibitem[\protect\citeauthoryear{{Vos}, {N{\'e}meth}, {Vu{\v c}kovi{\'c}},
  {{\O}stensen}  \& {Parsons}}{{Vos} et~al.}{2018}]{Vos2018a}
{Vos} J.,  {N{\'e}meth} P.,  {Vu{\v c}kovi{\'c}} M.,  {{\O}stensen} R.,
  {Parsons} S.,  2018, \mn@doi [\mnras] {10.1093/mnras/stx2198}, \href
  {http://adsabs.harvard.edu/abs/2018MNRAS.473..693V} {473, 693}

\bibitem[\protect\citeauthoryear{{Webbink}}{{Webbink}}{1984}]{Webbink1984}
{Webbink} R.~F.,  1984, \mn@doi [\apj] {10.1086/161701}, \href
  {http://adsabs.harvard.edu/abs/1984ApJ...277..355W} {277, 355}

\bibitem[\protect\citeauthoryear{{Werner}, {Deetjen}, {Dreizler}, {Nagel},
  {Rauch}  \& {Schuh}}{{Werner} et~al.}{2003}]{Werner2003}
{Werner} K.,  {Deetjen} J.~L.,  {Dreizler} S.,  {Nagel} T.,  {Rauch} T.,
  {Schuh} S.~L.,  2003, in {Hubeny} I.,  {Mihalas} D.,   {Werner} K.,  eds,
  ASPCS Vol. 288, Stellar Atmosphere Modeling. p.~31 (\mn@eprint {}
  {astro-ph/0209535})

\bibitem[\protect\citeauthoryear{{Zorotovic} \& {Schreiber}}{{Zorotovic} \&
  {Schreiber}}{2017}]{Zorotovic2017}
{Zorotovic} M.,  {Schreiber} M.~R.,  2017, \mn@doi [\mnras]
  {10.1093/mnrasl/slw236}, \href
  {http://adsabs.harvard.edu/abs/2017MNRAS.466L..63Z} {466, L63}

\bibitem[\protect\citeauthoryear{{Zorotovic}, {Schreiber}, {G{\"a}nsicke}  \&
  {Nebot G{\'o}mez-Mor{\'a}n}}{{Zorotovic} et~al.}{2010}]{Zorotovic2010}
{Zorotovic} M.,  {Schreiber} M.~R.,  {G{\"a}nsicke} B.~T.,   {Nebot
  G{\'o}mez-Mor{\'a}n} A.,  2010, \mn@doi [\aap] {10.1051/0004-6361/200913658},
  \href {http://adsabs.harvard.edu/abs/2010A%26A...520A..86Z} {520, A86}

\makeatother
\end{thebibliography}


\label{lastpage}
\end{document}